\documentclass[preprint,12pt]{elsarticle}

\usepackage{amssymb}
\usepackage{amsmath}
\usepackage{tabularx}
\usepackage{dcolumn}
\newcommand{\RN}[1]{%
  \textup{\uppercase\expandafter{\romannumeral#1}}%
}

\journal{Journal of Computational physics}

\begin{document}

\begin{frontmatter}



\title{A hybrid continuum surface force for the three-phase VOF method}


\author[a]{Chunheng Zhao}

\author[a]{Jacob Maarek}
\author[c,d]{Seyed Mohammadamin Taleghani}
\author[a,b]{Stephane Zaleski}

\address[a]{Sorbonne Universit\'{e} and CNRS, Institut Jean Le Rond d'Alembert UMR 7190, F-75005 Paris, France}
\address[b]{Institut Universitaire de France, Paris, France}
\address[c]{Universitat Politècnica de Catalunya (UPC), Barcelona School of Industrial Engineering (ETSEIB), Diagonal 647, Barcelona, Spain}
\address[d]{Université de Lorraine, Faculté des Sciences et Technologies, F-54500 Vandœuvre-lès-Nancy, France}

\begin{abstract}
We propose a hybrid continuum surface force (CSF) formulation to model the interface interaction within the three-phase volume of fluid (VOF) method. Instead of employing the height function globally, we compute the curvature based on a smooth fraction function near the region of the triple contact line. In addition, we apply the isotropic finite difference method to calculate derivatives, and the current scheme readily accommodates adaptive mesh refinement, greatly enhancing efficiency. We rigorously validate the hybrid CSF using two benchmark problems that have received limited attention in previous studies of the three-phase VOF method. Using the hybrid CSF method, we accurately predict the behavior of a liquid lens under specific surface tension ratios. Furthermore, the simulation results of the equilibrium morphology of two contacting droplets are consistent with the theoretical expectations.
\end{abstract}

\begin{graphicalabstract}

\end{graphicalabstract}

\begin{highlights}
\item A hybrid continuum surface force coupled with three-phase VOF method. 
\item Basic benchmark validations for three-phase VOF method.
\end{highlights}

\begin{keyword}
Three-phase VOF \sep Hybrid CSF
\PACS 0000 \sep 1111
\MSC 0000 \sep 1111
\end{keyword}

\end{frontmatter}


\section{Introduction}
Numerical modeling of multi-phase fluids, particularly three-phase fluids, holds significant utility in various industrial applications. Examples include its use in microfluidics~\cite{whitesides2006origins,pannacci2008equilibrium}, froth floatations~\cite{quintanilla2021modelling,shean2011review,zhao2023interaction}, and even 3D printing~\cite{shahrubudin2019overview,villegas2019liquid}. The modeling of three-phase fluids poses several primary challenges. One challenge is choosing a proper interface capturing method, and the other one is computing the surface forces near the triple contact lines, which necessitates the handling of singularities. In such cases, diffused interface methods, such as the Cahn-Hilliard method~\cite{cahn1958free,kim2005phase}, and the conservative phase-field method~\cite{chiu2011conservative,sun2007sharp,aihara2019multi}, prove valuable. These methods require the consideration of a finite interface region near the interface. However, when dealing with a multi-scale system, approaching the sharp interface limit using diffused interface methods can lead to expensive computation~\cite{yue2010sharp,zhao2023engulfment}. Concurrently, sharp interface methods, including Level-Set, Volume of Fluid (VOF), and Front-Tracking methods, can be applied to simulate three-phase flow~\cite{lie2005front,vu2013front,wang2004color,kromer2023efficient,mithun2018numerical}. The interface reconstruction of three phases with the VOF method is more challenging than for two phases due to intersecting planes,~\cite{kromer2023efficient}, leading to inaccuracies in the morphology near the triple contact lines. Consequently, when surface tension forces are computed by the continuum surface tension force (CSF)~\cite{brackbill1992continuum} where the curvature relies on the inaccurate fraction function, it results in incorrect physics. To address this issue, we propose a hybrid CSF based on the three-phase VOF method. We approximate a smooth fraction function near the triple contact lines and evaluate the curvature by the isotropic finite difference method with the smoothed fraction function. In regions away from the triple contact line, we compute the interfacial curvature using the height function method, as is done in general two-phase flow simulations~\cite{popinet2009accurate}. Previous studies have often lacked essential benchmark problems due to the challenges associated with implementing the three-phase VOF method. In our research, we address this gap by conducting two validations of this scheme, using benchmark tests related to liquid lenses~\cite{zhao2023engulfment,kalantarpour2020three}, as well as the equilibrium morphology of two contacting droplets~\cite{zhao2023interaction,wang2020modelling}. Our simulation results exhibit a high degree of consistency with the theoretical solutions.

\section{Methodology}
\subsection{Surface force modeling}
\begin{figure*}[hbt!]
  \centering
  \includegraphics[width=\textwidth]{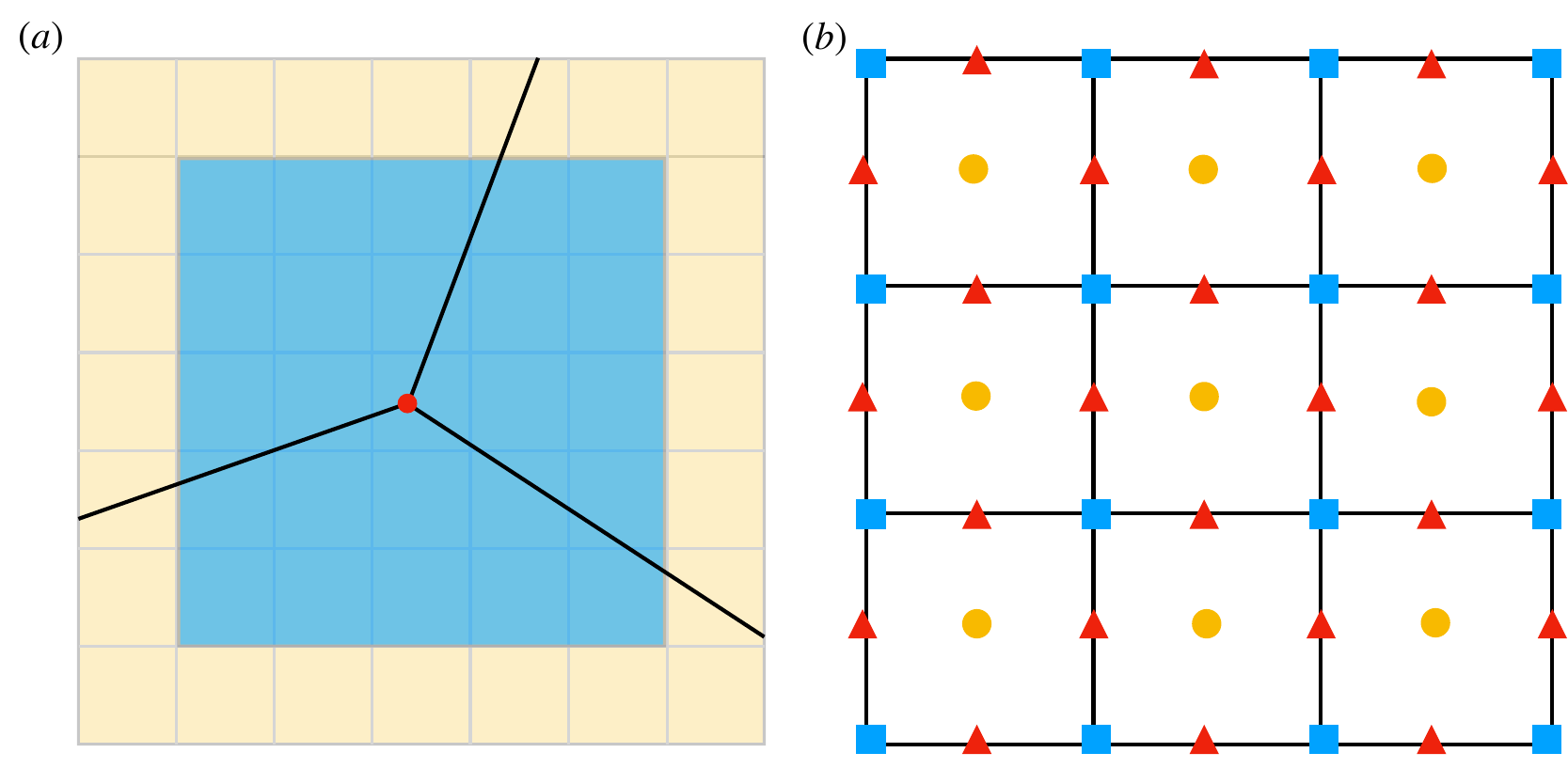}
  \caption{\label{node} (a) Region of the triple contact line (indicated by a red dot) is marked in blue. The curvature is calculated from the smooth fraction in the blue region. (b) Schematic of the staggering type: centered: marked by yellow dots; face: marked by red triangles; vertex: marked by blue squares. }
\end{figure*}
The following derivation of the hybrid CSF method is based on the 2-D problem, and the entire approach can be easily accommodated to the 3-D version. We start from the CSF for the general two-phase flow:
\begin{equation}\label{csf}
    \boldsymbol{F}_s=\sigma\kappa \delta_s \boldsymbol{n},
\end{equation}
with $\kappa$ denotes the interface curvature, $\delta_s$ is the interface delta Dirac function, $\boldsymbol{n}$ is the interface normal vector, $\sigma$ is the surface tension between two fluids. In the VOF, the above Eq.~\ref{csf} can be modified as a function of the fraction $f$:
\begin{equation}\label{modified}
    \boldsymbol{F}_s=\sigma\kappa \nabla f.
\end{equation}
Based on the two-phase CSF, the three-phase CSF can be obtained from the modified surface tensions for each phase. Consider an n-phase flow with phases noted fluid $a$, $a=1-n$. The surface tension between a pair of fluids $a, b$ is noted $\sigma_{ab}$. The modified surface tensions can be defined as $\sigma_a=0.5(\sigma_{ab}+\sigma_{ac}-\sigma_{bc})$, and the CSF in the three-phase VOF method can be interpolated by three surface tension forces:
\begin{equation}\label{sum}
    \boldsymbol{F}_s=\sum_{a=1}^{3}\sigma_a\kappa_a\nabla f_a,
\end{equation}
where $\kappa_a$ and $f_a$ represent the curvature and the fraction for fluid $a$, respectively. Normally, the gradient of the fraction can be directly computed by the central finite difference method for each phase, and the curvature is evaluated by the height function. When combined with a well-balanced CSF method~\cite{popinet2009accurate}, it is able to effectively eliminate the well-known spurious currents for two-phase flow simulation. For three-phase flow, the advection, and reconstruction of the fraction function are not accurate. We find that the largest problem affecting capillary-driven three-phase flows is the computation of the net force at the triple point caused by erroneous curvature evaluations given from the height function method. In this case, the following strategy is applied for the curvature computation near the triple contact lines.

As shown in the figure.~\ref{node}(a), we separate the region with interfaces into two parts, one is the $5\times5$ window, where the center is the triple contact line, and the other one is the two-phase interface region. In the triple contact line region, the curvature evaluation is based on a smooth function located at the lattice center proposed in~\cite{liu2022computation}, where the new smooth fraction function is given as:
\begin{equation}\label{2Dsmo}
\begin{split}
        \bar{f}^c_{i,j}=\frac{4}{9}f^c_{i,j}+&\frac{1}{9}(f^c_{i+1,j}+f^c_{i-1,j}+f^c_{i,j+1}+f^c_{i,j-1})+\\
        &\frac{1}{36}(f^c_{i+1,j+1}+f^c_{i+1,j-1}+f^c_{i-1,j+1}+f^c_{i-1,j-1}).
\end{split}
\end{equation}
Here the superscript $c$ represents the centered staggering type. In addition, we have faced, and vertex staggering types which are shown in figure~\ref{node}(b). The subscripts $i$ and $j$ indicate the horizontal and vertical positions.
The gradient of the smooth fraction function is evaluated by the isotropic finite difference:
\begin{equation}\label{2Diso}
    \left(\partial_x \bar{f}\right)^c_{i,j}\approx\frac{1}{3\Delta x}\left(\bar{f}^c_{i+1,j}-\bar{f}^c_{i-1,j}\right)+\frac{1}{12\Delta x}\left(\bar{f}^c_{i+1,j+1}-\bar{f}^c_{i-1,j+1}+\bar{f}^c_{i+1,j-1}-\bar{f}^c_{i-1,j-1}\right),
\end{equation}
after which the curvature located on the center can be computed as $\kappa^c=\nabla\cdot \left[\frac{(\nabla\bar{f})^c}{|(\nabla \bar{f})^c|}\right]$. Finally, we are able to compute the curvature at the face by the mean value $\kappa^f_{i+0.5,j}=0.5(\kappa^c_{i,j}+\kappa^c_{i+1,j})$, and the surface tension force can be computed as:
\begin{equation}\label{final}
    \boldsymbol{F}_s=\sum_{a=1}^{3}\sigma_a\kappa_a^f(\nabla f_a)^f.
\end{equation}
It is noted that the last term $(\nabla f)^f$ is evaluated by central finite difference where 
\begin{equation}
    (\partial_x f)^f_{i+0.5,j}\approx\frac{1}{\Delta x}(f^c_{i+1,j}-f^c_{i,j}).
\end{equation}

A geometric VOF advection scheme is used for the transport of the interface. Each phase is advected according to a directionally split VOF advection scheme that conserves mass for each phase separately ~\cite{WEYMOUTH20102853}. Due to inconsistent reconstruction of each phase near the triple point, it is not guaranteed after advection that the volume fractions add to one. To remediate this problem, we normalize the volume fractions after advection. The process of re-normalization results in the loss of global mass conservation properties. 
Results demonstrate that mass loss is reduced as the grid is refined. We leave the improvement of the three-phase advection scheme, either by exact reconstruction in three-phase cells or a modification of the phasic flux computation, as the topic of a future study.  

The variable-density incompressible Navier-Stokes equations are solved with the centered solver native to Basilisk. The methods used in the solver have been validated extensively in capillary-driven interfacial flows ~\cite{POPINET2015336,popinet2009accurate}.

Wavelet-based adaptive mesh refinement for tree-based grids is used to accelerate the runtime of the simulations performed. The scheme used is described and visualized in ~\cite{POPINET2015336, VANHOOFT}. In addition to the common phase fractions and velocity field commonly used as refinement criteria, we added a constraint that forces maximum refinement in a $5\times5$ stencil around the triple point. This constraint minimizes mass loss and ensures an optimal calculation for curvatures near the triple point. As the triple point is represented by a point in 2-D or a line in 3-D the added cost of this additional constraint is negligible.

\section{Simulation Results}
In the following simulations, to eliminate the effect of the density ratio and the viscosity ratio, we only consider the same density $\rho$, and viscosity $\mu$ for three different fluids. 

\subsection{Liquid lens problem}
\begin{figure*}[hbt!]
  \centering
  \includegraphics[width=\textwidth]{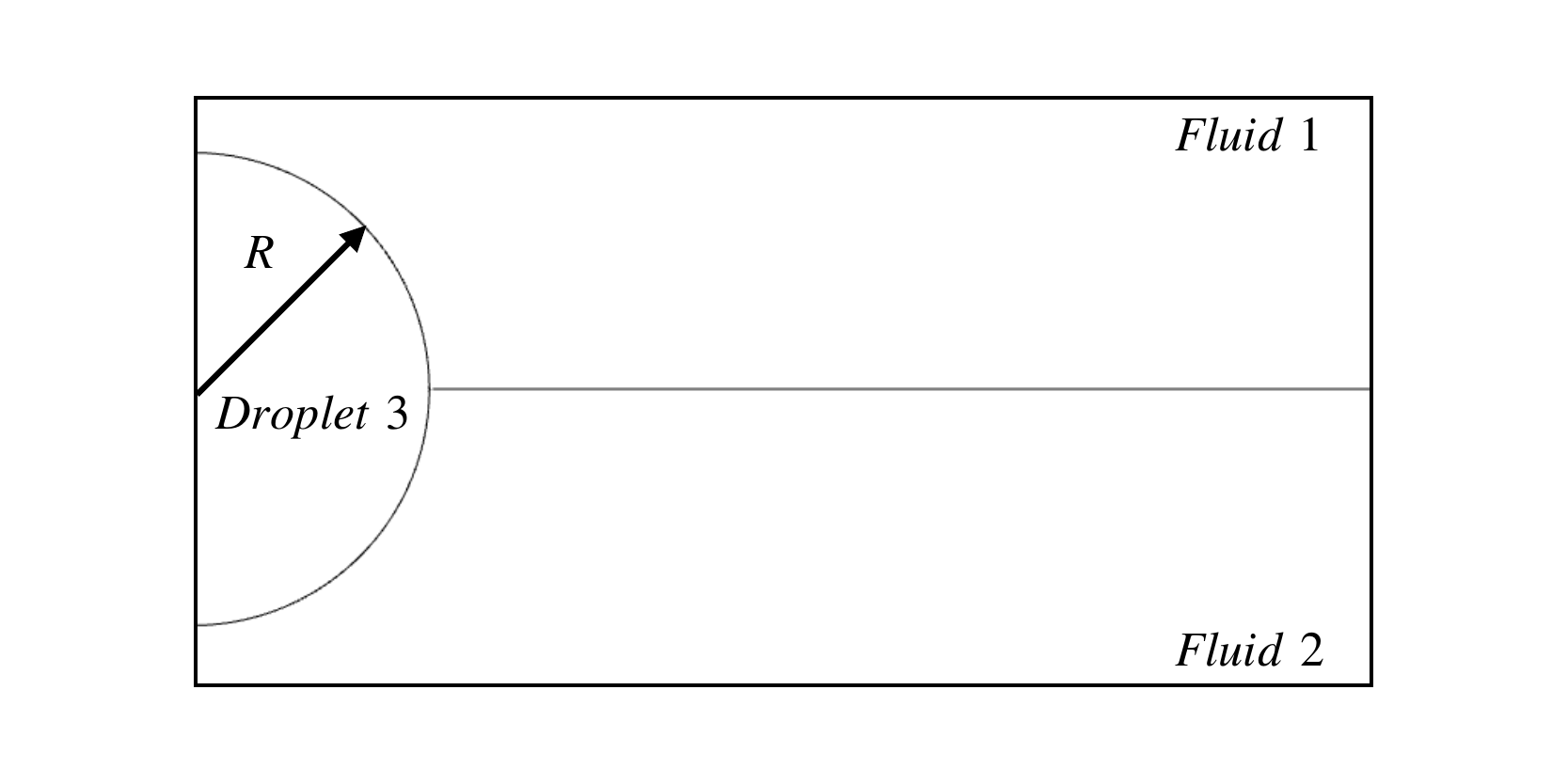}
  \caption{\label{liqinit}Initialization of the liquid lens simulation.}
\end{figure*}
As depicted in Figure~\ref{liqinit}, we initialize a single droplet with radius $R$ between two fluids. Over time, the triple contact line evolves until it reaches an equilibrium state where the three surface tensions are balanced.
\begin{figure*}[hbt!]
  \centering
  \includegraphics[width=\textwidth]{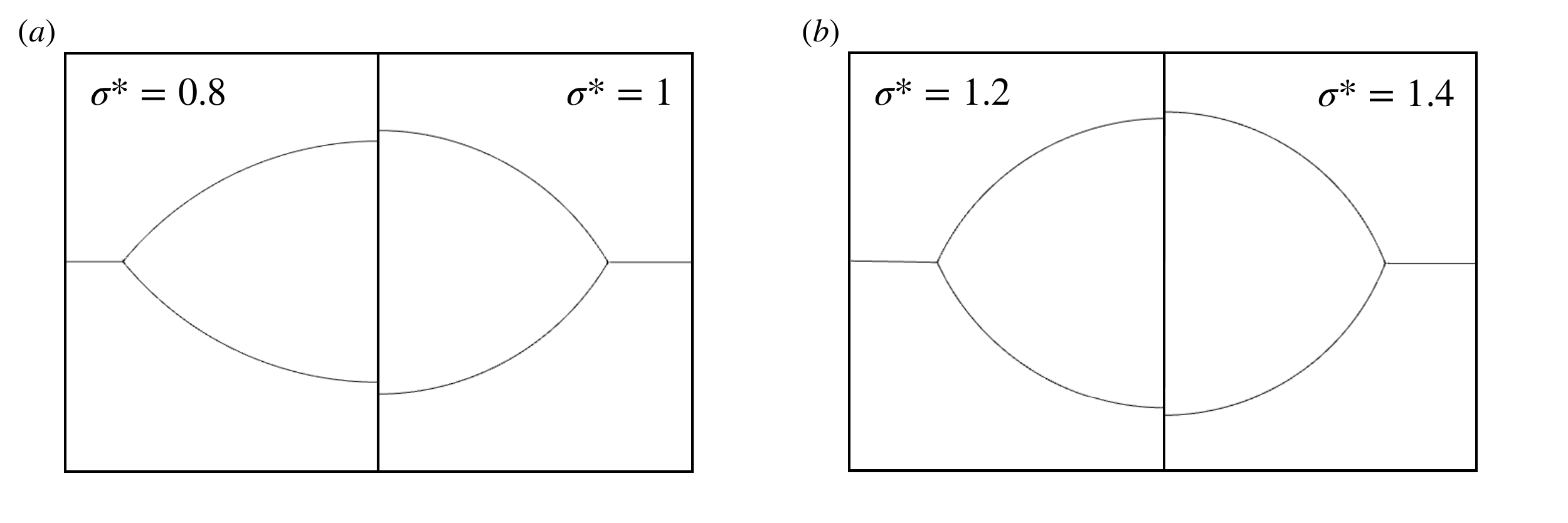}
  \caption{\label{liquidlens}Equilibrium morphology of the liquid lens simulations with various surface tensions. (a) $\sigma^*=0.8$ (left panel) and $\sigma^*=1.0$ (right panel). (b) $\sigma^*=1.2$ (left panel) and $\sigma^*=1.4$ (right panel).}
\end{figure*}
Theoretically, as we modify the ratio of three surface tensions $\sigma_{12}, \sigma_{13},\sigma_{23}$ between fluid 1 and fluid 2, fluid 1 and droplet 3, fluid 2 and droplet 3, the equilibrium contact angle $\theta$ for each phase can be obtained by the relation equation:

\begin{equation}
      \frac{\sin\theta_1}{\sigma_{23}}=\frac{\sin\theta_2}{\sigma_{13}}=\frac{\sin\theta_3}{\sigma_{12}}.
\end{equation}
The contact angle for each fluid can be derived as:
\begin{equation}
  \theta_a=\cos^{-1}\left(-\frac{\sigma_{ab}^2 +\sigma_{ac}^2 -\sigma_{bc}^2}{2\sigma_{ab}\sigma_{ac}}\right).
\end{equation}
The parameter setup is as follows: we initialize the surface tension ratios with different values, specifically $\sigma_{12}:\sigma_{13}:\sigma_{23}=(1:0.8:0.8), (1:1:1), (1:1.2:1.2),$ and $(1:1.4:1.4)$. We represent the ratio between $\sigma_{13}$ and $\sigma_{12}$ as $\sigma^*$. The simulation results are obtained at $T=100T_\mu$ to ensure that the system reaches an equilibrium state. Here, $T_\mu=\mu R/\sigma_{12}$ represents the viscous time scale. The theoretical solution for the length $d$ between the triple contact line and the initial center of the droplet can be expressed as follows:
\begin{table*}
\caption{The analytic solutions versus the simulation results for liquid lens length.}
\begin{tabular}{lllll}

  $\sigma^*$   &    $Level-7$       & $Level-8$     & $Level-9$      & analytic solution\\ \hline

$\sigma^*=0.8$  & 0.585938  &0.597656    & 0.607422 &0.6128\\

$\sigma^*=1.0$  & 0.523438  &0.535156    & 0.548828 &0.5540\\

$\sigma^*=1.2$  & 0.492188  &0.511719    & 0.517578 &0.5220\\

$\sigma^*=1.4$  & 0.476562  &0.488281    & 0.494141 &0.5014

\end{tabular}\label{liquid}
\end{table*}

\begin{equation}
\frac{1}{d^2} =\frac{1}{8A}\left(\frac{2(\pi-\theta_1)-\sin(2(\pi-\theta_1)))}{\sin^2(\pi-\theta_1)}
+\frac{2(\pi-\theta_2)-\sin(2(\pi-\theta_2)))}{\sin^2(\pi-\theta_2)}\right),
\end{equation}
where $A=\pi R^2$ is the area of the initial droplet. Table~\ref{liquid} presents the ratio of the liquid lens length to the initial radius, denoted as $0.4d/R$, obtained at various resolutions ranging from $\Delta x=1/64$ (level-7) to $\Delta x=1/256$ (level-9). In the last column, we compare our simulation results with the corresponding analytic solutions. Notably, as we increase the resolution, our simulation results approach the analytic solutions.

\subsection{Morphology diagram}
\begin{table}
\caption{Parameters table of the equilibrium morphology simulation.}

\begin{tabular}{cccccccc}
 Case Number&$\sigma_{23}$&$\sigma_{13}/\sigma_{23}$&$\sigma_{12}/\sigma_{23}$&$S_1$&$S_2$&$S_3$\\ \hline
 $\RN{1}$-A(1)  & 0.01  & 0.5  & 1.7 &$<0$&$<0$ &$>0$ \\

 $\RN{1}$-A(2)  & 0.01  & 1    & 2.2 &$<0$&$<0$ &$>0$ \\

 $\RN{1}$-B(1)  & 0.01  & 1.7 & 0.5  &$<0$ &$>0$ &$<0$\\

 $\RN{1}$-B(2)  & 0.01  & 2.2 & 1    &$<0$&$>0$ &$<0$ \\

 $\RN{2}$      & 0.01  & 0.3 & 0.3  &$>0$&$<0$ &$<0$ \\

 $\RN{3}$(1)  & 0.01  & 1    & 1    &$<0$&$<0$ &$<0$ \\

 $\RN{3}$(2)  & 0.01  & 0.5  & 1    &$<0$&$<0$ &$<0$ \\

 $\RN{3}$(3)  & 0.01  & 1    & 0.5  &$<0$&$<0$ &$<0$ \\

 $\RN{3}$(4)  & 0.01  & 1    & 1.5  &$<0$&$<0$ &$<0$\\

 $\RN{3}$(6)  & 0.01& 100  & 100    &$<0$&$<0$ &$<0$ \\

\end{tabular}

\end{table}\label{diagramt}
\begin{figure*}[hbt!]
  \centering
   \includegraphics[width=\textwidth]{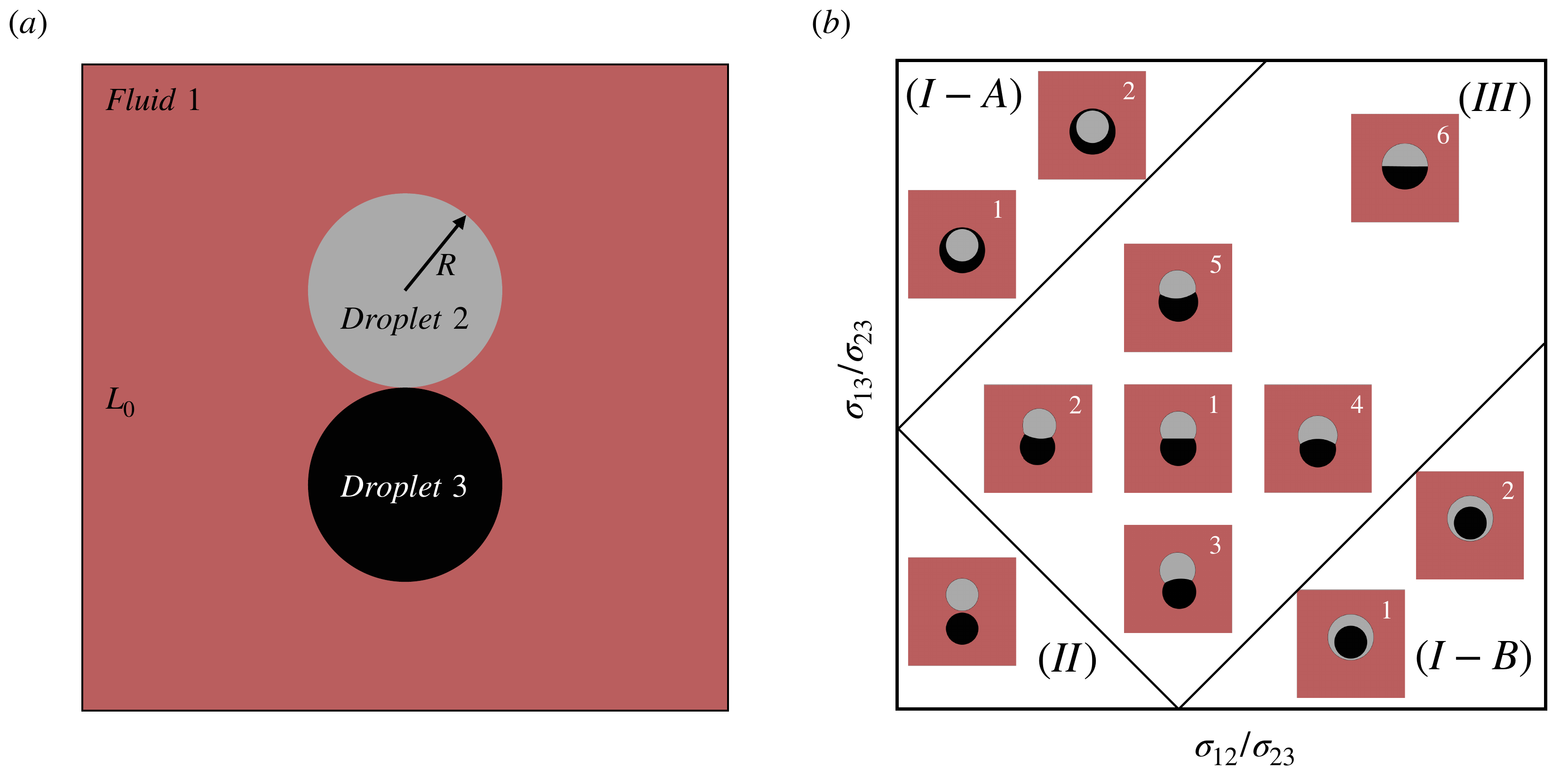}
  \caption{\label{diagram}(a) Initialization of the morphology simulation of two contacting droplets. (b) Morphology diagram of the equilibrium shapes. Regions ($\RN{1}$-A), ($\RN{1}$-B) denote the double emulsion regime, Region ($\RN{2}$) is the separate morphology regime, and Region ($\RN{3}$) is the partially engulfed regime.}
\end{figure*}
The second test aims to apply this scheme to simulate the equilibrium morphology diagram for a specific combination of surface tensions.

\begin{figure*}[hbt!]
  \centering
  \includegraphics[width=\textwidth]{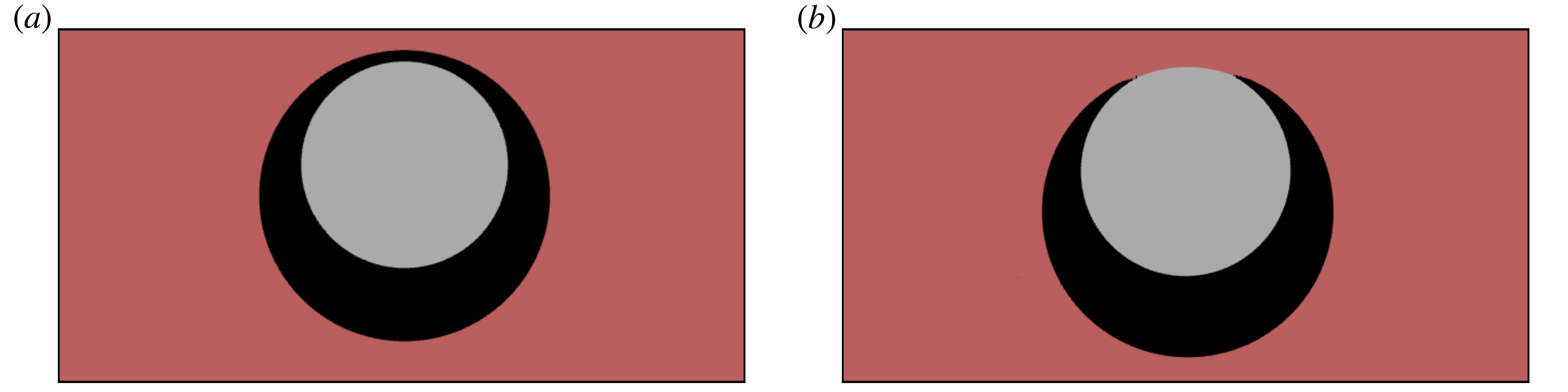}
  \caption{\label{comp}Equilibrium morphology of contacting droplets with surface tension ratio $\sigma_{12}:\sigma_{13}:\sigma_{23}=1.7:0.5:1$ when the curvature near the triple contact line is evaluated by (a) smooth fraction $\&$ isotropic finite difference and (b) height function.}
\end{figure*}

As depicted in figure~\ref{diagram}(a), we initiate two droplets of equal size, each with a radius of $R=0.15L_0$, in contact with each other at the center of a square domain, where $L_0=2$ represents the side length of the square domain. The entire process is driven by the interplay of surface tensions among the three phases, and the specific surface tension values between different components are provided in Table.~\ref{diagramt}. We define the following surface tensions: $\sigma_{23}$ between the two droplets, $\sigma_{12}$ between the background fluid 1 and droplet 2, and $\sigma_{13}$ between fluid 1 and droplet 3. After establishing these surface tension definitions, we derive the spreading factors as follows: $S_1=\sigma_{23}-\sigma_{12}-\sigma_{13}$, $S_2=\sigma_{13}-\sigma_{23}-\sigma_{12}$, and $S_3=\sigma_{12}-\sigma_{23}-\sigma_{13}$. In theory, when $S_2>0$ or $S_3>0$, the droplet will eventually become fully engulfed, with the outer component being either droplet 2 or droplet 3, respectively. Conversely, when $S_1>0$, the two droplets will separate. Therefore, as illustrated in Table~\ref{diagram}, the occurrence of a double emulsion, where one droplet is entirely engulfed by another, is expected in Region $\RN{1}$. A separated droplet morphology will manifest in Region $\RN{2}$, while a partially engulfed morphology is anticipated in Region $\RN{3}$.

In order to achieve the equilibrium state, our simulations run for an extended duration, typically exceeding $T > 200T_\mu$, where $T_\mu = \mu R/\sigma_{23}$. Subsequently, we present the resulting morphologies graphically in Figure~\ref{diagram}(b) for a more intuitive visualization. 

It is noteworthy that our simulation results exhibit a high degree of consistency with theoretical solutions. 
\begin{figure*}[hbt!]
  \centering
  \includegraphics[width=\textwidth]{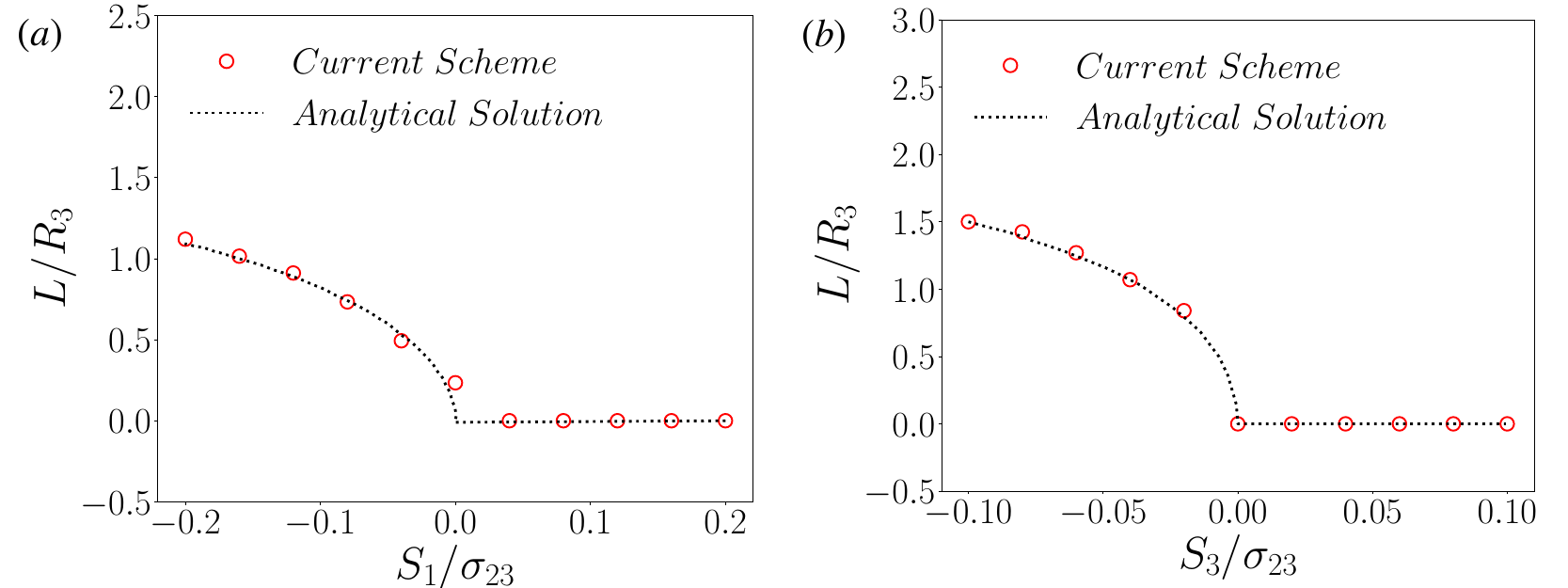}
  \caption{\label{length}Final distance $L$ between two triple contact lines with various spreading factors. (a) Region ($\RN{2}$) $\&$ Region($\RN{3}$). (b) Region ($\RN{1}$) $\&$ Region($\RN{3}$).}
\end{figure*}
We provide a brief comparison between the hybrid CSF method and the previous approach in Figure~\ref{comp}. In this test, the positive spreading factor of droplet 3 ($S_3>0$) indicates that droplet 2 will completely engulf droplet 3. Figure~\ref{comp}(b) displays the simulation result obtained using the previous method, which fails to replicate the fully engulfed morphology. In contrast, the new surface force model correctly captures the formation of a double emulsion (see figure~\ref{comp}(a)).

Further tests are conducted to investigate morphology transformations under various spreading factors close to critical value $S\sim0$. Initially, we gradually increased the spreading factor of fluid 1 $S_1$ by adjusting the surface tensions $\sigma_{12}$ and $\sigma_{13}$ within the range of $[0.004,0.006]$, while keeping $\sigma_{23}$ fixed at 0.01. The results in Figure~\ref{length}(a) illustrate that the distance between two triple contact lines was accurately simulated. Upon further increasing the spreading factor, a separated morphology shown in Region ($\RN{2}$) becomes evident.

Conversely, in the transitional region between Region $\RN{1}$ and Region $\RN{3}$, we modified the initial radius of droplet 3, setting it to $0.2L$. We maintained $\sigma_{13}$ and $\sigma_{23}$ at 0.01, while adjusting $\sigma_{12}$ within the range of $[0.019,0.021]$. The outcomes, as depicted in Figure~\ref{length}(b), consistently align with the analytical solution.

\section{Concluding remarks}
In this study, we presented a strategy of surface tension modeling in the three-phase VOF method. Instead of globally implementing the height function for curvature evaluation, we employed a regular isotropic finite difference to compute the curvature based on a smooth fraction function near the triple contact line. By using this scheme, the promotion of the prediction of the morphology evolution is obvious.

Through the liquid lens simulation, we were able to obtain convergence results to the analytic solution as we increased the mesh resolution. A comparison between using the current method and the height function was conducted on a comprehensive modeling of the equilibrium morphology when two droplets contact. The predictions of the current method were consistent with the theoretical analysis. Meanwhile, the height function failed to simulate the correct equilibrium morphology in this test. The proposed method accurately computed the net capillary force near the triple contact lines allowing us to model capillary-driven three-phase flows. 

It is worth noting that the advection method for three-phase flow modeling can be highly improved. We would like to further investigate the advection method of the three-phase VOF method as shown in recent studies~\cite{kromer2023efficient}.
 \bibliographystyle{main} 
 \bibliography{main}

\end{document}